\documentclass[prl,aps,twocolumn,showpacs,superscriptaddress]{revtex4-1}
\usepackage{epsfig}
\usepackage{bm,hyperref}
\usepackage{amssymb}
\usepackage{mathrsfs}
\usepackage{amsmath}

\begin{document}

\title{Intermittent Josephson effect with feedback voltage and temperature oscillations in graphite-coated nanocapsules with superconducting TaC core}

\author{Dianyu Geng}
\author{Zhenhua Wang}
\email{zhwang@imr.ac.cn}
\author{Da Li}
\author{Zhidong Zhang}
\affiliation{Shenyang National Laboratory for Materials Science, Institute of Metal Research, and International Center for Materials Physics,	 Chinese Academy of Sciences, Shenyang 110016, People's Republic of China}
\author{Xiaolin Wang}
\affiliation{Institute for Superconducting and Electronic Materials, Faculty of Engineering, Australian Institute for Innovative Materials, University of Wollongong, Wollongong, Australia}

\begin{abstract}
An intermittent Josephson effect in the form of voltage and temperature oscillations in the voltage - current curves near 2 K is observed in pellets consisting of superconducting TaC nanocapsules coated with graphite. This phenomenon is attributed to non-equilibrium conditions, when Cooper pairs across a junction, which stimulate the emission of photons and the feedback temperature change of the junction. It occurs in a three-dimensional granular framework composed of TaC/carbon/TaC tunneling junctions with a Mott metal-insulator transition, below the critical temperature T$_c$ of non-ideal type-II superconductor TaC.
\end{abstract}

\pacs{75.20.-z,75.75.Lf,61.46.Hk}

\maketitle

The Josephson tunneling effect in superconductors is one of the most important quantum effects \cite{01jose, 02jose,03jean}. Nanocapsules (or core-shell nanostructures) have already revealed abundant new physics and potential applications \cite{04sait,05zhang}. Superconducting nanocapsules consisting of superconductors (of any type) as core and normal (i.e. non-superconducting) materials as shells are particularly interesting in terms of the fabrication of junctions. The behavior of tunneling supercurrents in Josephson junctions may be very unlike ordinary experience. For instance, Josephson predicted non-equilibrium properties of coupled superconductors\cite{02jose}, such as the occurrence of dc supercurrents (i.e. the transfer of Cooper pairs across the barrier), provided the potential difference is such that energy can be conserved by absorption or stimulated emission of a photon (or multiphoton). Intermittent-type chaos occurring in rf- and dc-current-driven Josephson junctions was also investigated \cite{06ben, 07kautz, 08goldhirsch, 09octavio,zumofen10}.

It is particularly important to detect the transformation between Cooper pairs and electrons during non-equilibrium processes in Josephson junctions. These inspire us to fabricate Josephson junctions using carbon-coated TaC nanocapsules and to investigate their superconductive characteristics related to the junction tunneling effect. It is expected that a collective variant of the single-electron tunneling and/or supercurrent effects may take place in a three-dimensional framework of granular superconductors with tunneling junctions \cite{li11,wang12,wang13}. Although there have been many studies on intermittent-type Josephson effect \cite{06ben, 07kautz, 08goldhirsch, 09octavio,zumofen10}, up to our knowledge, there has been no report on intermittent Josephson effects with feed-back temperature change on materials. In this Letter, we report the structure and electric, magnetic, and superconducting properties of TaC(C) nanocapsules. Near 2 K, voltage and temperature oscillations are observed in the current-voltage curves, which are attributed to that electron pairs intermittently tunneling through the Josephson non-zero voltage junctions emit photons \cite{02jose}. At low temperature, the superconductive currents first take place in the junctions. emitting photons, arising temperature and resistance, and then the superconductive currents decrease (or stop), decreasing the Cooper pairs and the temperature until the next cycle (the supercurrents take place again at temperature low enough). There is a feedback of the temperature change and the voltage - current function of the junctions. The combination effect of the superconductor TaC and the Mott metal-insulator transition of graphite are likely responsible for the oscillations.

The carbon-coated TaC nanocapsules studied in this work have been fabricated via an arc discharge process in ethanol vapor and argon atmosphere. The details of this method have been described elsewhere \cite{li11,wang12,wang13}. A piece of pure Ta metal and a graphite rod were used as anode and cathode, respectively. The phase and structure of the TaC(C) nanocapsules were investigated by X-ray diffraction (XRD) using Cu-K$_{\alpha}$ ($\lambda$ = 0.15405 nm) radiation. The particle size was analyzed by high-resolution transmission electron microscopy (TEM). The transport and magnetic properties were measured in a Quantum Design superconducting quantum interference device (SQUID) magnetometer, with maximum temperature tolerance of less than 5 mK below 5 K. The relative error of the temperature and voltage data is less than 0.1\%. For the resistivity measurements, as-prepared TaC(C) nanocapsules were pressed into pellets under 1.2 GPa. I-V curves were measured with a Keithley 2400 Source Meter and a Keithley 2182 Nanovoltmeter. The temperature dependence of the resistivity and the I-V curves were measured using the SQUID in the sweep mode.

\begin{figure}
 \includegraphics[width=7cm]{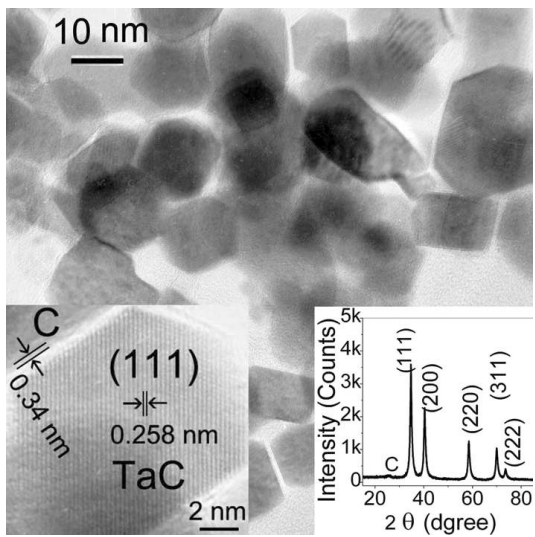}
 \caption{TEM morphology and core/shell structure of TaC(C) nanocapsules. The left inset shows a crystal grain of TaC(C), and the right inset contains the XRD pattern.}\label{fig1}
\end{figure}

The TEM image in Fig. \ref{fig1} shows the core/shell structure of the TaC(C) nanocapsules. The crystal plane spacings of the cores for some of the nanocapsules in Fig. \ref{fig1} are only in one direction, and they usually consist of only one TaC grain coated with a carbon shell. The left inset of Fig. \ref{fig1} shows a single nanocapsule which contains a single grain of TaC as core and carbon as shell. The characteristic spacing of the strong-reflection (111) plane of TaC is 0.258 nm. The size of the nanocapsules ranges between 10 and 20 nm. The XRD pattern shown in the right inset of Fig. \ref{fig1} indicates that the phases in the nanocapsules are indeed TaC and C. The thickness of the spacing in the carbon shell is 0.34 nm, which indicates that the shell is graphite with (002) orientation. The TaC grains are separated by the graphite, which acts as a conductance barrier between two TaC grains. The electrons are trapped inside isolated TaC grains. The grain connections in the system of nanocapsules may be considered as zero-dimensional, although the grain sizes in three dimensions are in nanometers. The tunneling framework consists of superconductor-insulator-superconductor ($\ldots$ TaC/(C)/TaC$ \ldots$) tunneling junctions, which vary among themselves because of the different thicknesses of the graphite shells and the different sizes of the TaC core.

\begin{figure}
\includegraphics[width=11cm]{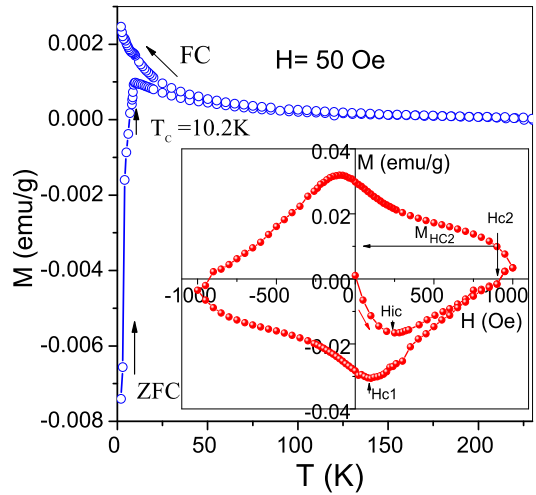}
\caption{Zero-field-cooled (ZFC) and field-cooled (FC) magnetization of TaC(C) nanocapsules as a function of temperature, measured in a magnetic field of 50 Oe. The inset shows a hysteresis loop at 2 K, which is typical for a non-ideal type-II superconductor.
}\label{fig2}
\end{figure}

The temperature dependence of the zero-field-cooled (ZFC) and field-cooled (FC) magnetization of TaC(C) nanocapsules, measured in a magnetic field of 50 Oe, which is lower than the first critical field, is presented in Fig. \ref{fig2}. The ZFC curve shows a clear superconductive transition temperature T$_c$ of 10.2 K from paramagnetism to diamagnetism, at the point where the TaC cores become superconducting. The FC curve initially exhibits Pauli paramagnetism and indicates strong magnetic vortex pinning features. The inset in Fig. \ref{fig2} shows the magnetic hysteresis loop at 2 K, which is asymmetric because TaC(C) is a non-ideal type-II superconductor with T$_c$ = 10.2 K \cite{shig14,toth15,fink16,yosi17}, and there is in the magnetic fields from 0 Oe to 1000 Oe, and then to -1000 Oe to 1000 Oe. On the initial magnetization curve, at the initial critical field H$_{ic}$ = 250 Oe, the negative magnetization begins to increase. On the loop, at H = 100 Oe, the magnetization begins to shift. H$_{c1}$ = 100 Oe is the lower critical field, and H$_{c2}$$\approx$920 Oe is the upper critical field, showing that the superconductivity of the TaC(C) nanocapsules is typical of a non-ideal type-II superconductor, for which the magnetization process is irreversible. For ideal type-II superconductors, the remanent magnetization at H$_{c2}$ is zero, and the magnetic flux lattice is symmetric \cite{yosi17}. For the TaC(C) nanocapsule conductor in the first quadrant for H = 920 Oe, M$_{Hc2}$ is not zero, but approximately 0.01 emu/g. (H is from 1000 Oe to 920 Oe.) The magnetic hysteresis loop changes with the magnetic field, and the magnetic flux lattice is not symmetrical in the TaC(C) nanocapsules. For the upper critical field, the higher the magnetic-field range, the higher the M$_{Hc2}$ is. It trends toward a limiting value which is smaller than the remanent magnetization. This property is useful for obtaining a high critical current density in superconductors \cite{bean18}.

\begin{figure}
\includegraphics[width=11cm]{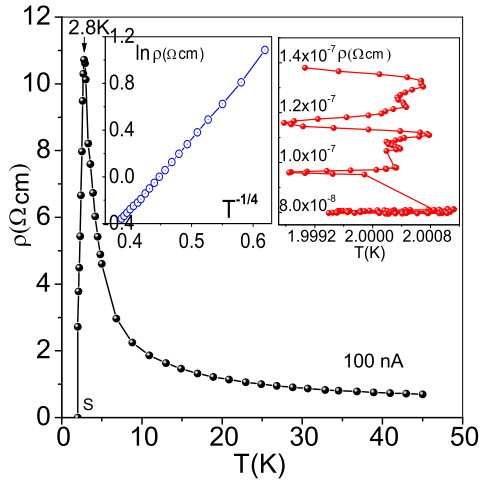}
\caption{Temperature dependence of the resistivity ($\rho$) of a pellet of TaC(C) nanocapsules measured at a current of 100 nA. The right inset presents a magnification of ¦Ñ at the lowest temperatures (point S). In the left inset, ¦Ñ is plotted on a logarithmic scale against T$^{-1/4}$.}\label{fig3}
\end{figure}

Graphite is a semiconductor with zero activation energy at zero temperature \cite{wall19} and a Mott insulator with a metal-insulator transition \cite{case20,mott21}. Figure 3 presents the temperature dependence of the resistivity ($\rho$) measured at zero field at a current of 100 nA by means of the four-probe dc method. The $\rho$-T curve shows a metal-insulator transition, which can be ascribed to the graphite shells of the TaC nanocapsules. Above 2.8 K, the resistivity is semiconductor-like, and below 2.8 K, it is metallic and precipitous decreased. At 2 K the resistivity of the nanocapsules has reached zero, which means that the TaC nanocrystals are in the superconducting state. The right inset shows a magnification of the behavior of $\rho$ near 2 K. The left inset of Fig. \ref{fig3} shows a plot of $\rho$ on a logarithmic scale against T$^{-1/4}$. It illustrates that the resistivity is that of a Mott insulator, which is similar to what has been reported for NbC(C)-C \cite{li11}, carbon-coated Sn nanoparticles \cite{wang12}, and Mo carbide nanoparticles embedded in a carbon matrix \cite{wang13}. This explains the electrical-conduction mechanism of the TaC(C) nanocapsule composite, which contains an intrinsic framework of nanocrystalline TaC and graphite. The main process in the resistivity is a Mott metal-insulator transition \cite{case20,mott21}.

Figure 4 shows V-I curves that were subsequently measured at 11, 9, 7, 5, 3, and 2 K by means of the two-probe dc method (foil: length 3.34 mm; width, 1.88 mm; thickness, 0.06 mm) in the current range from 300 $\mu$A to 0 $\mu$A, with decreasing steps of 2.5 $\mu$A. During the measurement process the time from 300 $\mu$A to 0 $\mu$A was 1800 s. The top scale of time(s) corresponds to the time for every curve measurement. At 11 K, which is above T$_c$, the V-I curve follows Ohmic behavior, with V/I = R = constant. Below T$_c$, single-electron tunneling takes place with conductance close to 4e$^2$/h. Comparing the curves at 9 K, 7 K, 5 K, and 3 K, the voltage increases with decreasing temperature, and the curves gradually become curved, indicating that the conduction mechanism starts deviating from Ohm's law.
\begin{figure}
\includegraphics[width=11cm]{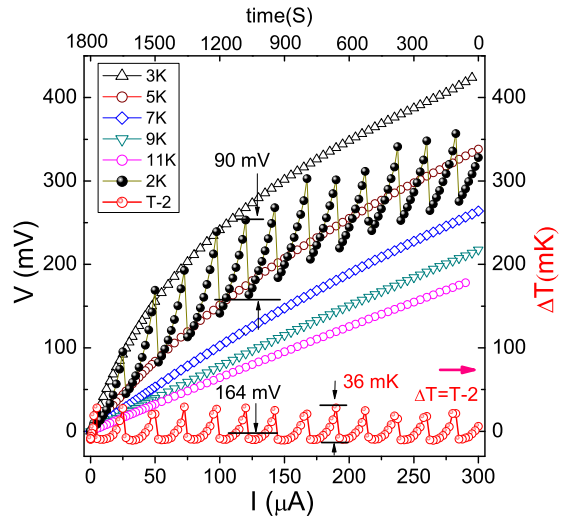}
\caption{Current (I) versus Voltage (V) and temperature fluctuations ($\triangle$ T= T-2). V-I curves of a pellet of TaC(C) nanocapsules successively measured at 11, 9, 7, 5, 3, and 2 K. The lowest curve presents the temperature fluctuations  $\triangle$ T(= T-2) of V-I curve measured at 2 K. The top scale of time(s) corresponds to the time for every curve. The voltage oscillation, 90 mV = $V_{s}=\Sigma^{i=n}_{i=1}V_{si}$, as described below in the text, results from the supercurrent at that moment; combined voltage from the Ohmic part and the single-electron tunneling part is 164 mV. The maximum temperature difference is 36 mK for the I-$\triangle$T curve.}\label{fig4}
\end{figure}

In Fig. \ref{fig3}, on the $\rho$-T curve, from 10 K to 3 K, the higher the temperature, the higher $\rho$ is, which corresponds to what is shown in Fig. \ref{fig4} that the lower the temperature, the higher the voltage is. Very interestingly, at 2 K, the voltage exhibits oscillatory behavior as a function of current, which can not arise from Ohmic current and single-electron tunneling.

During measurement in the sweep mode at 2 K, the temperature of the sample was observed to oscillate despite the temperature control, because when H = 0 there are still some remaining magnetic fluxes inside. To indicate how the temperature oscillates with the current, an I-$\Delta$T ($\Delta$T = T - 2 K) curve is presented at the bottom of Fig. \ref{fig4}. In the oscillations, the voltage and temperature increase over a short time ($<$ 0.05 s) and then decrease slowly ($>$ 120 s). The oscillations in temperature and voltage are synchronous.

From Figs.\ref{fig2} and \ref{fig3}, we know that at 2 K, the TaC nanograins are superconductors. The Cooper pairs tunneling at non-zero voltage through the TaC/C/TaC junctions give rise to the emission of photons, which results in an increase in the temperature. In fact, the voltage exhibits oscillatory behavior accompanied by temperature change ($\Delta$T = T-2). Figure 5 represents that the voltage (V) and $\Delta$T versus time (t) curves, which were successively measured at 2 K, with a current I = 50 nA at the absence of a magnetic field (H = 0). The measurement method is shown as Fig. \ref{fig4}. The V-t and $\Delta$T-t curves oscillate like a music score, and the big oscillation amplitude is 0.16 mV. The resistance, V/I, does not follow Ohmic and single tunneling electron behavior and the voltage jumps with  T, which indeed exhibits a quantum effect.
\begin{figure}
\includegraphics[width=10cm]{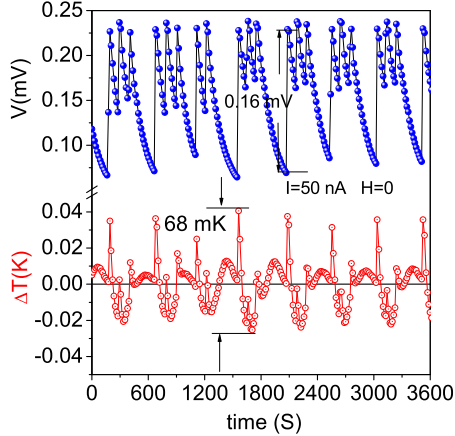}
\caption{Voltage (V) versus time (t) and temperature fluctuations ( T =$\Delta$T-2) curves of a pellet of TaC(C) nanocapsules successively measured at 2 K, with H=0 and I=50 nA. The V-t and $\Delta$T-t curves oscillate like a music score, and the big oscillation amplitude is 0.16 mV. The changing range for $\Delta$T is 68 mK. V and $\Delta$T also exhibit as functions of time and the characteristic quantum indeterminacy.}
\label{fig5}
\end{figure}

In Fig. \ref{fig2}, T$_c$ is observed to be 10.2 K, and in Fig. \ref{fig3}, the maximum resistance temperature is 2.8 K. In superconductors, the free energy, which corresponds to the condensation energy, equals E $\propto$ (T$_c$-T)$^2$ \cite{anne22}. For the supercurrent through TaC/C/TaC tunneling junctions, the temperature difference needs to be at least over T$_c$-T $\approx$ 10.2K - 2.8 K = 7.4K to overcome the barriers between TaC and graphite.

The supercurrent, the electric Ohmic current, the single-electron tunneling current, and their corresponding voltages coexist. To explain the oscillations, all TaC nanocapsules are assumed to be isolated from each other. The TaC(C) nanocapsules are pressed into a specimen, which forms a three-dimensional granular framework composed of TaC/C/TaC tunneling junctions. We assume the number of tunnel junctions in the pellet to be equal to n and use the equivalent lumped circuit model \cite{mccu23,hans24} for the experiments.

For the i$^{th}$ tunneling junction, the total current I$_i$ is composed of the supercurrent, the electric Ohmic current, and the single-electron tunneling current. The total current measured in the I-V curves is
\begin{equation}
\sum^n_{i=1} I_i = I \nonumber.
\end{equation}

\begin{figure}
\includegraphics[width=10cm]{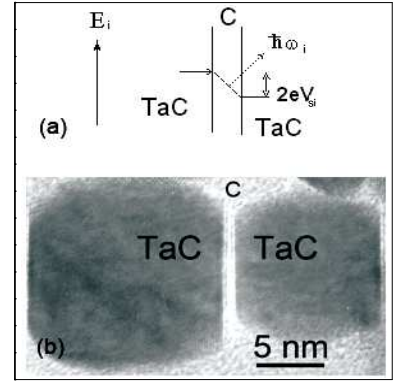}
\caption{(a) A Cooper pair tunnels through the $i^{th}$ Josephson junction, emitting a photon, ${\hbar\omega_i=2eV_{si}}$  , as it crosses the interface between the two TaC(C) nanocapsules. E is the free energy in the TaC nanocapsules.  (b) TEM morphology of one Josephson junction in the TaC (C) nanocapsules, and the thickness of graphite layers between the two TaC grains is 1 nm.}\label{fig6}
\end{figure}

In the system for the equivalent model, the i$^{th}$ tunneling junction may be written as follows \cite{02jose}:
\begin{equation}
\hbar \omega_i = 2 e V_{si} \nonumber ,
\end{equation}
where $\omega_i$ is the frequency of the emitted photon when a Cooper pair is tunneling through the i$^{th}$ TaC/C/TaC tunneling junction; V$_{si}$ is the non-zero potential difference between the two sides of the i$^{th}$ junction (see Fig. \ref{fig6}). Then:
\begin{equation}
V_s = \sum^n_{i=1} V_{si} = \frac{\hbar}{2e} \left( \sum^n_{i=1} \omega_i \right ) (i = 1, 2, 3, \ldots, n),
\end{equation}
where V$_s$ is the voltage barrier that must be surmounted for Cooper pairs to cross the junctions and emit photons. There are involved the special rules for the interfering amplitudes for the Cooper pairs and electrons that occur in the processes. The amplitude for the Bose particles is the sum of the two interfering amplitudes, and for the Fermi ones it is minus.

When measured the I-V curves some Cooper pairs are interference. For the states with Bose particles, the probability in the Bose case is larger than that of the particles acted independently \cite{anne22,Feyn25}. When the tunneling effects take place, the probability amplitudes of Cooper pairs in graphite abruptly increase and give rise to the emission of photons and consequently, the temperature jumps up; increasing the resistance of graphite. Then the supercurrent stops and the number of Copper pairs is quickly decreased, which are transformed to single electrons. Subsequently, the temperature begins to drop down. With decreasing the temperature of the system, the number of Cooper pairs is gradually increased, and the resistance of graphite decreases until the next cycle. The probability amplitudes of Copper pairs also are a function of the time (See Fig. \ref{fig5}) and exhibited quantum indeterminacy. In fact, there is a competitive mechanism between the junction barrier and the free energy of the superconducting carriers in the TaC nanograins. A phase transition occurs during the supercurrent goes through the junctions, where a boson (Cooper pair) transforms into two fermions (electrons), and this process stimulates the emission of a photon. After photons are emitted, the temperature of the TaC/C/TaC nanocapsules rises abruptly; the voltage then jumps, and the supercurrent stops because the resistance has been quickly risen (See Fig. \ref{fig3} from 2 K to 2.8 K). The probability amplitudes of Cooper pairs are driven by the particle statistics of Cooper pairs, in which the thermodynamic observable has an abrupt change, leading to a jump of voltage, a quantum behavior \cite{anne22,Feyn25}. The graphite barrier is a Mott insulator with a metal-insulator transition. As the temperature gradually decreases, the barrier is reduced. However, in nano-TaC grain superconductors the free energy is increased as the temperature falls. The barrier is decreased and the free energy is increased, until the free energy overcomes the barrier, and the another cycle begins again. The temperature and the voltage are gradually reduced, which is driven by electron interactions. The Mott metal-insulator transition of the graphite shells promotes oscillations, as is shown in Fig. \ref{fig6}.

Therefore, the tunneling mechanisms of the Cooper pairs in the non-ideal type-¢ò superconductor TaC and the Mott metal-insulator transition in graphite would be responsible for the voltage and temperature oscillations in the nanocapsules. As shown in Fig. \ref{fig4}, while this occurs, $V_s$ = $\sum_{i=1}^{n}$$V_{si}$ = 90 mV. The combined voltage of the Ohmic part and the single-electron tunneling part is 164 mV.

In summary, oscillations in voltage and temperature have been observed in TaC(C) nanocapsules. The Josephson intermittent effect occurs in a non-equilibrium system of ($\ldots$TaC/(C)/TaC$\ldots$) tunneling junctions, with a function of the time. The mechanism is that the supercurrents go through the Josephson junctions and emit photons with the feedback resistance,temperature change of the TaC(C) junctions.

The authors acknowledge the support of the National Natural Science Foundation of China under grant number 51102244 and the National Basic Research Program (No. 2012CB933103), the Ministry of Science and Technology of China. This work is also partly supported by the Australian Research Council through a Discovery project.


\begin{thebibliography}{25}
\bibitem{01jose} {B. D. Josephson, Phys. Lett. {\bf 1}, 251 (1962).}
\bibitem{02jose} {B. D. Josephson, Rev. Mod. Phys. {\bf 36}, 216 (1964).}
\bibitem{03jean} {B. Jeanneret and S. P. Benz, Eur. Phys. J. Special Topics {\bf 172}, 181 (2009).}
\bibitem{04sait} {Y. Saito, Carbon {\bf 33}, 979 (1995).}
\bibitem{05zhang} {Z. D. Zhang, in: H.S. Nalwa (Ed.), Encyclopedia of Nanoscience and Nanotechnology,Vol. {\bf 6}, American Scientific, California, (2004) 77-160.}
\bibitem{06ben} E. Ben-Jacob, I. Goldhirsch, Y. Imry and S. Fishman, Phys. Rev. Lett. {\bf 49}, 1599 (1982).
\bibitem{07kautz} R. L. Kautz, Rep. Prog. Phys. {\bf 59}, 935-992  (1996).
\bibitem{08goldhirsch} I. Goldhirsch Y. Imry, G. Wasserman and E. Ben-Jacob Phys. Rev. B. {\bf 29}, 1218 (1984).
\bibitem{09octavio} M. Octavio, Phys. Rev. B {\bf 29}, 1231 (1984).
\bibitem{zumofen10} G. Zumofen and J. Klafter, Phys. Rev. E {\bf 47}, 851 (1993).
\bibitem{li11} {D. Li, W.F. Li, S. Ma, and Z.D. Zhang, Phys. Rev. B {\bf 73}, 193402 (2006).}
\bibitem{wang12} {Z. H. Wang, D. Y. Geng, Z. Han, and Z. D. Zhang, J. Appl. Phys. {\bf 108}, 013903 (2010).}
\bibitem{wang13} {Z. H. Wang, D. Li, D. Y. Geng, S. Ma, W. Liu, and Z. D. Zhang, J. Mater. Res. {\bf 24}, 2229 (2009).}
\bibitem{shig14} {T. Shigaki, S. M. Oh, J.G. Li, and D. W. Park, Sci. Technol. Adv. Mater. {\bf 6} (2005) 111-118.}
\bibitem{toth15} {L. E. Toth, Transition Metal Carbides and Nitrides, Academic Press: New York, (1971).}
\bibitem{fink16} {H. J. Fink, A. C. Thoresen, E. Parker, V. F. Aackay, and L. Toth, Phys. Rev. {\bf 138}, A1170 (1965).}
\bibitem{yosi17} {Y. Yosida and I. Oguro, Physica C {\bf 434}, 173 (2006).}
\bibitem{bean18} {C.P. Bean, Phys. Rev. Lett. {\bf 8}, 250 (1962).}
\bibitem{wall19} {P. R. Wallace, Phys. Rev. {\bf 71}, 622 (1947).}
\bibitem{case20} {A. Casey, H. Patel, J. Ny\'eki, B. P. Cowan, and J. Saunders, Phys. Rev. Lett. {\bf 93}, 115301 (2003).}
\bibitem{mott21} {N. F. Mott, Rev. Mod. Phys. {\bf 40}, 677 (1968).}
\bibitem{anne22} {J. F. Annett, Superconductivity, Superfluids and Condensates, Oxford University Press, (2004) 6, 73.}
\bibitem{mccu23} {D.E. McCumber, J. Appl. Phys. {\bf 39}, 2503 (1968).}
\bibitem{hans24} {P.K. Hansma, G.I. Rochlin, and J.N. Sweet, Phys. Rev. B. {\bf 4}, 3003 (1971).}
\bibitem{Feyn25} {R. P. Feynman, R. B. Leighton, M. Sands, The Feynman Lectures on Physics {\bf 3}, Printed in the United States of America, (1964).}
\end{thebibliography}
\end{document}